\documentclass[12pt,epsf]{article}
\usepackage{amssymb,amsmath}
\usepackage{graphicx}
\newcommand{\be}{\begin{equation}}
\newcommand{\ee}{\end{equation}}
\newcommand{\bea}{\begin{eqnarray}}
\newcommand{\eea}{\end{eqnarray}}
\newcommand{\beas}{\begin{eqnarray*}}
\newcommand{\eeas}{\end{eqnarray*}}
\newcommand{\ba}{\begin{array}}
\newcommand{\ea}{\end{array}}

\newcommand{\tr}{\mathrm{Tr}}

\newcommand{\nbox}{{\,\lower0.9pt\vbox{\hrule \hbox{\vrule height 0.2 cm \hskip 0.19 cm \vrule height 0.2 cm}\hrule}\,}}

\def\href#1#2{#2}

\begin{document}
\begin{titlepage}
\hfill
\vbox{
    \halign{#\hfil         \cr
           hep-th/0606014  \cr
           } % end of \halign
      }  % end of \vbox
\vspace*{20mm}
\begin{center}
{\Large \bf Little String Theory from a \\ Double-Scaled Matrix Model}

\vspace*{15mm}
\vspace*{1mm}

{Henry Ling, Ali Reza Mohazab, Hsien-Hang Shieh, Greg van Anders, \\ Mark Van Raamsdonk}

\vspace*{1cm}

{Department of Physics and Astronomy,
University of British Columbia\\
6224 Agricultural Road,
Vancouver, B.C., V6T 1W9, Canada}

\vspace*{1cm}
%%\maketitle
\end{center}

\begin{abstract}

Following Lin and Maldacena, we find exact supergravity solutions dual to a class of vacua of the plane wave matrix model by solving an electrostatics problem. These are asymptotically near-horizon D0-brane solutions with a throat associated with NS5-brane degrees of freedom. We determine the precise limit required to decouple the asymptotic geometry and leave an infinite throat solution found earlier by Lin and Maldacena, dual to Little String Theory on $S^5$. By matching parameters with the gauge theory, we find that this corresponds to a double scaling limit of the plane wave matrix model in which $N \to \infty$ and the 't Hooft coupling $\lambda$ scales as $\ln^4(N)$, which we speculate allows all terms in the genus expansion to contribute even at infinite $N$. Thus, the double-scaled matrix quantum mechanics gives a Lagrangian description of Little String Theory on $S^5$, or equivalently a ten-dimensional string theory with linear dilaton background. 

\end{abstract}

\end{titlepage}

\vskip 1cm
\section{Introduction}

Type IIA Little String Theory \cite{lst1} describes the degrees of freedom of NS5-branes in type IIA string theory in a decoupling limit where the string coupling $g_s$ is taken to zero keeping $\alpha'$ fixed (i.e. focusing on energies of order $(\alpha')^{-1/2}$). What remains is believed to be a six-dimensional interacting non-gravitational theory with a Hagedorn density of states. In the infrared, the theory flows to the interacting (0,2) conformal field theory, but in general the theory does not have the properties of a local quantum field theory (in particular, it is believed to be nonlocal). For a review of Little String Theory, see \cite{lst}. 

Since there is no direct Lagrangian description of the theory (though a DLCQ formulation \cite{dlcq} and a description via deconstruction \cite{dcon} have been proposed), the main tool for analyzing Little String Theory has been its gravity dual, the near-horizon NS5-brane solution of type IIA string theory, given for large $r$ where the IIA picture is valid by 
\beas
ds^2 &=& N_5 \alpha'(-dt^2 + d\vec{x_5}^2 + dr^2 + d \Omega_3^2) \cr
e^{\phi} &=& g_s e^{-r} 
\eeas
with $N_5$ units of $H$ flux through the $S^3$. Even in this description, the theory is difficult to work with, since the background contains a linear dilaton, sending the theory to strong coupling (M-theory) in the infrared region of the geometry. 

Recently, Lin and Maldacena \cite{lm} have found a related supergravity solution in which the flat five-dimensional part of the geometry corresponding to the spatial NS5-brane worldvolume directions has been replaced by an $S^5$. We reproduce the (somewhat complicated) full supergravity solution in appendix A, but for large radius, the metric and dilaton become simply
\beas
ds^{2} &=& N_5 \alpha ^{\prime }\left[ 2r \left(
-dt^{2}+d\Omega _{5}^{2}\right) +dr ^{2}+d \Omega_3^2\right]  \notag \\
e^{\Phi } &=& g_{s} e^{-r } \; ,
\eeas
again with $N_5$ units of $H$ flux through the $S^3$. We see that the new solution retains the linear dilaton behavior and constant volume $S^3$ permeated by H-flux, so it is natural to associate this solution with NS5-branes on $S^5$. However in this case, the full solution has a tunable maximum value for the dilaton at $r=0$ and a tunable maximum curvature, so we have a regime where supergravity is everywhere valid.\footnote{Unfortunately, the solution contains Ramond-Ramond fields, so string theory is difficult.} \footnote{A similar situation occurs in \cite{gk, gk2}, though in the present case, more of the R-symmetry is preserved. See also \cite{mnun} and \cite{mnast} for discussions of the type IIB Little String Theory compactified on $S^2$ and $S^3$ respectively.}

The main goal of this paper will be to explicitly describe a field theory dual for string theory on this solution, and thus a Lagrangian field theory definition of Little String Theory on $S^5$.

The broader context for our story is a D0-brane quantum mechanics analogue of the model of Polchinski and Strassler \cite{ps} for D3-branes. The field theory we consider is the plane-wave matrix model \cite{bmn}, a mass-deformation of the maximally supersymmetric D0-brane quantum mechanics. In our case, the mass deformation is maximally supersymmetric, preserving 32 supercharges including an $SU(2|4)$ symmetry \cite{dsv2,kp,kpark}. The theory has a discrete spectrum, a dimensionless parameter that acts as a tunable coupling constant \cite{dsv1}, and a large number of degenerate supersymmetric vacua preserving the $SU(2|4)$ supersymmetry. 

Before the mass-deformation, the D0-brane quantum mechanics is dual to string theory on the near-horizon D0-brane solution of supergravity \cite{imsy}. As with the near-horizon NS5-brane solution discussed above, this becomes strongly coupled in the infrared region of the geometry, and we must go to an eleven-dimensional description. However, the mass-deformation provides an infrared cutoff for the theory, so we might expect that solutions dual to the various vacua of the plane-wave matrix model (for large enough mass) would have an everywhere valid IIA description. 

This picture was verified recently by Lin and Maldacena \cite{lm}. Following \cite{llm}, they searched for type IIA supergravity solutions with $SU(4|2)$ symmetry, and showed that solutions of type IIA supergravity corresponding to all vacua of the plane-wave matrix model could be constructed in terms of the solutions to a class of axially symmetric electrostatics problems involving charged conducting disks in 3 dimensions. While they did not solve the electrostatics problem except in certain limiting cases, they showed that solutions of this type generally contain throats with non-contractible 3-spheres carrying H-flux. These regions of the geometry may be associated with fivebrane degrees of freedom described by the matrix model.

In this paper, we focus on the simplest class of vacua of the plane-wave matrix model, for which the dual geometry has only a single NS5-brane throat. We solve the appropriate electrostatics problem to find an exact supergravity solution, and determine the precise limit of this solution (which depends on three parameters) needed to decouple the throat region, yielding the explicit infinite-throat solution of Lin and Maldacena corresponding to Little String Theory on $S^5$. By understanding how the parameters of the supergravity solution match up with the parameters of the gauge theory, we then see what this limit corresponds to in the matrix model. We find that the corresponding limit is a limit of large $N$ with the 't Hooft coupling also taken to infinity in a particular way, roughly $\lambda \sim \log^4(N)$, while focusing on the excitations around a specific vacuum of the matrix model (the one corresponding to our original supergravity solution). 

Since we are taking a strict large-$N$ limit in the field theory, we might naively expect that the corresponding gravity dual should be a free string theory. This is true asymptotically, due to the linear dilaton background, but not for finite values of $r$, so there should still be a genus expansion on the string theory side. We conjecture that this is reproduced in the gauge theory in a way very similar to the double-scaling limits used to describe low-dimensional string theories in terms of matrix models.\footnote{This conclusion was predicted by Herman Verlinde.} In our case, the 't Hooft coupling is scaled towards a critical coupling $\lambda_c = \infty$ in such a way that the various terms in the matrix model genus expansion all contribute despite $N$ being infinite. Assuming this picture is correct, we are able to make predictions for the large $\lambda$ behavior of the full set of genus $n$ diagrams in perturbation theory (section 6.1).

The paper is organized as follows. In section 2, we review various aspects of the plane-wave matrix model, including decoupling limits that have been discussed in the past. In section 3, we review the Lin-Maldacena ansatz for gravity duals to the matrix model vacua and the electrostatics problems that need to be solved in order to find the solutions. We then provide an exact solution for the simplest such problem, which requires determining the potential due to parallel charged conducting disks in a specified background potential. In section 4, we discuss the matching of parameters between the gravity solutions and the matrix model. In section 5, we determine how to scale the parameters in our solution to obtain the Lin-Maldacena solution for Little String Theory on $S^5$, and then use the correspondence with the gauge theory to determine the matrix model description of Little String Theory on $S^5$. We also discuss the limit that gives a solution dual to the maximally supersymmetric theory of D2-branes on $S^2$, and the gravity interpretation of the 't Hooft limit of the matrix model. In section 6, we discuss the results. In 6.1, we explore the consequences of our result that the Little String Theory on $S^5$ is obtained as a double scaling limit of the matrix model. In section 6.2, we describe an infinite-parameter family of supergravity solutions similar to (and with the same symmetries as) the Lin-Maldacena solution, and speculate on the matrix model description of these. Finally in section 6.3, we describe an application of our results to calculating energies of near-BPS states in the geometry. Various appendices provide more technical results.  

\section{Gauge Theory} \label{gaugesec}

The plane-wave matrix model is described by a Hamiltonian \cite{bmn}\footnote{Here, all quantities are dimensionless. Appendix \ref{mapp} gives the relation between these conventions and the usual matrix theory conventions.}
\bea
\label{PPmatrix2}
H &=& \tr \left( {1 \over 2} P_A^2 + {1 \over 2} (X_i/3)^2 + {1 \over 2} (X_a/6)^2
+ {i \over 8} \Psi^\top \gamma^{123} \Psi \right. \cr
&& \qquad \left.+  {i \over 3} g \epsilon^{ijk} X_i X_j X_k - {g \over 2} \Psi^\top \gamma^A [X_A, \Psi] - {g^2 \over 4} [X_A, X_B]^2 \right)
\eea
where $A=1, \dots , 9$, $i=1, \dots ,3$, and $a=4, \dots ,9$. Here, the scalars $X_A$ and 16-component fermions $\Psi$ are hermitian $N \times N$ matrices, and $P_A$ is the matrix of canonically conjugate momenta. In addition, we have a Gauss law constraint that requires physical states to be invariant under the $U(N)$ symmetry transformation that act on the matrices as $M \to UMU^{-1}$. The model has one discrete dimensionless parameter, $N$, and a continuous dimensionless parameter $g$.

The set of classical vacua for the model are described by $X^a=0$, $X^i = {1 \over 3 g} J^i$, where $J^i$ give any $N$ dimensional representation of the $SU(2)$ algebra
\[
[J^i, J^j] = i \epsilon^{ijk} J^k \; .
\]  
These vacua are in one-to-one correspondence with partitions of $N$, since we may have in general $n_k$ copies of the $k$-dimensional irreducible representation such that $\sum_k k n_k = N$. 

This model has several interesting large $N$ limits (distinguished by which combination of $g$ and $N$ we hold fixed), which we now describe.

\subsubsection*{The M-theory limit}

According to the Matrix Theory conjecture \cite{bfss}, in the limit
\[
N \to \infty \qquad \qquad g^2/N^3 \; \; {\rm fixed} \; ,
\]
this model should describe M-theory on the maximally supersymmetric plane-wave background of eleven-dimensional supergravity,\footnote{This background has $ds^2 = ds^2_ {flat} + ({\mu^2 \over 9} x^i x^i + {\mu^2 \over 36} x^a x^a ) dx^+ dx^+$ and 
$F_{123+} = \mu$} with 
\[
\mu p^+ l_p^2 = {N \over g^{2 \over 3}} \qquad \qquad p^- / \mu = H  \; .
\]
Note that the quantities on the left are boost-invariant and dimensionless. 

States of M-theory on the plane-wave with zero light-cone energy are BPS configurations involving concentric spherical membranes and/or concentric spherical fivebranes. These correspond to vacua of the plane-wave matrix model \cite{bmn, msv}. 

We will be particularly interested in vacua involving only one type of irreducible representation, say $N_2$ copies of the $N_5$-dimensional irreducible representation,
\be
\label{ourvac}
X^i = {1 \over 3g} \left( \begin{array}{ccc} J^i_{N_5} & & \{ N_2 \; copies \} \cr & \ddots & \cr & & J^i_{N_5} \end{array} \right)  \; .
\ee 
In the M-theory large $N$ limit with $N_2$ fixed, this configuration describes the state with $N_2$ coincident M2-branes. This is plausible, since classically, such a configuration corresponds to $N_2$ coincident fuzzy spheres. On the other hand, if we keep $N_5$ fixed in the limit, taking $N_2$ to infinity, the configuration gives us $N_5$ coincident spherical M5-branes.\footnote{Evidence for this comes from the fact that the BPS excitations about this vacuum match the expected BPS excitations of coincident spherical M5-branes \cite{msv}.}   

In \cite{msv} it was pointed out that there are other interesting large $N$ limits that do not describe all of the degrees of freedom of M-theory, but rather focus in on degrees of freedom associated with the spherical branes.

\subsubsection*{The D2-brane limit}

To understand the next limit, consider the excitations around the vacuum (\ref{ourvac}) at finite $N$. Classically, this configuration corresponds to $N_2$ coincident (two-dimensional) fuzzy spheres, on which the modes have maximum angular momentum $N_5$. Fluctuations about this configurations are then described by noncommutative gauge theory on a fuzzy sphere with gauge group $U(N_2)$.   

If we compare this action arising from the matrix model with an action written as a noncommutative field theory with noncommutativity parameter $\theta$ and coupling $g_2$ on a sphere of radius $r$, we find that the field theory parameters are related to the matrix model parameters via
\[
{\theta \over r^2} = {1 \over N_5} \qquad \qquad g^2_2 r  = {g^2 \over N_5} \qquad \qquad Er = H \; .
\]
Thus, by taking 
\be
\label{d2lim}
N \to \infty \qquad \qquad g^2/N \; \; {\rm fixed} \qquad \qquad N_2 \; \; {\rm fixed} \; ,
\ee
we obtain a commutative field theory on a sphere. This field theory, written down in \cite{msv,lm} is essentially the low-energy theory of D2-branes, with mass terms for the scalars and fermions and a coupling of the radial scalar to the magnetic field, such that the whole theory preserves $SU(2|4)$ supersymmetry. Note that we end up with a D2-brane theory instead of an M2-brane theory since the limit (\ref{d2lim}) does not decompactify the M-theory circle.

We can also try to find a similar limit to describe decoupled fivebrane degrees of freedom:

\subsubsection*{The 't Hooft limit}

For small values of $g$ (with fixed $N$), we can study excitations about the various vacuum states perturbatively. In \cite{dsv1}, it was found that the parameter controlling perturbation theory is different for different vacua. For the vacuum with $N_2$ copies of the $N_5$-dimensional irreducible representation, perturbation theory is controlled by the combination $g^2 N_2$, so for example, the vacuum with $N_2=1, N_5=N$ is more weakly coupled than the vacuum with $N_5=1,N_2=N$. 

In particular, in the limit
\[
N \to \infty \qquad \qquad g^2 N \; \; {\rm fixed}  \; ,
\]
the coupling associated with the fivebrane vacua (with fixed $N_5$ with $N_2 \to \infty$) remains finite, while the coupling associated with the membrane vacua (fixed $N_2$ with $N_5 \to \infty$) or any other generic vacuum goes to zero. 

Again, this limit does not decompactify the M-theory circle, so it was suggested in \cite{msv} (also based on supergravity arguments) that this limit describes NS5-branes on a sphere. Below, we will see that it is a somewhat modified limit that is dual to the Lin-Maldacena gravity solution for NS5-branes on $S^5$. 

\section{Gravity} \label{gsec}

In the previous section, we have described the plane-wave matrix model, and various interesting large-$N$ limits. At finite $N$, we can think of the matrix model as a massive deformation of the maximally supersymmetric quantum mechanics describing low-energy D0-branes in flat-space, similar in spirit to the deformation of ${\cal N}=4$ SYM considered by Polchinski and Strassler \cite{ps}. The gravity dual of the undeformed theory is string theory on the near-horizon D0-brane geometry, so we expect that the gravity dual for the plane wave matrix model should be some infrared modification of this. 

Recently, Lin and Maldacena \cite{lm} (following \cite{llm}) searched for type IIA supergravity solutions preserving the same $SU(2|4)$ symmetry as the vacua of the plane-wave matrix model. Using an ansatz with this symmetry (reproduced in appendix \ref{sgapp}), they were able to reduce the problem of finding supergravity solutions to the problem of finding axially-symmetric solutions to the three-dimensional Laplace equation, with boundary conditions involving parallel charged conducting disks and a specified background potential. 

\subsubsection*{The electrostatics problem}

Common to all vacua, we have in the electrostatics problem an infinite conducting plate at $z=0$ (on which we may assume that the potential vanishes), and a background potential\footnote{This potential results by taking point charges $\mp Z^4/3$ at $r=0,z= \pm Z$ in a constant electric field $\vec{E} = -{2 \over 3} Z^2 \hat{z}$ in the limit $Z \to \infty$.}  
\be
\label{vinf}
V_\infty = V_0(r^2 z - {2 \over 3} z^3) \; .
\ee
In addition, corresponding to a matrix model vacuum with $Q_i$ copies of the $d_i$-dimensional irreducible representation, we have conducting disks with charge $Q_i$ parallel to the infinite plate and centered at $r=0$, $z = d_i$. In order that the supergravity solution is non-singular, the radii $R_i$ of the disks must be chosen so that the charge density at the edge vanishes.\footnote{To see that this should be possible, note that without a background field, the charge density on a conducting disk diverges as an inverse square root near the edge. On the other hand, we have an inward electric field coming from the background potential that increases linearly with the radius of the disk. Thus, for a large enough disk, the tendency for the charge on the disk to bunch up at the edge should be balanced by the action of the inward electric field so that the charge density vanishes.} 

Thus, for each vacuum of the plane-wave matrix model, we have an electrostatics problem, whose solution (a potential $V(r,z)$) feeds into the equations (\ref{IIA ansatz last}) to give a supergravity solution.

\begin{figure}
\centering
\includegraphics{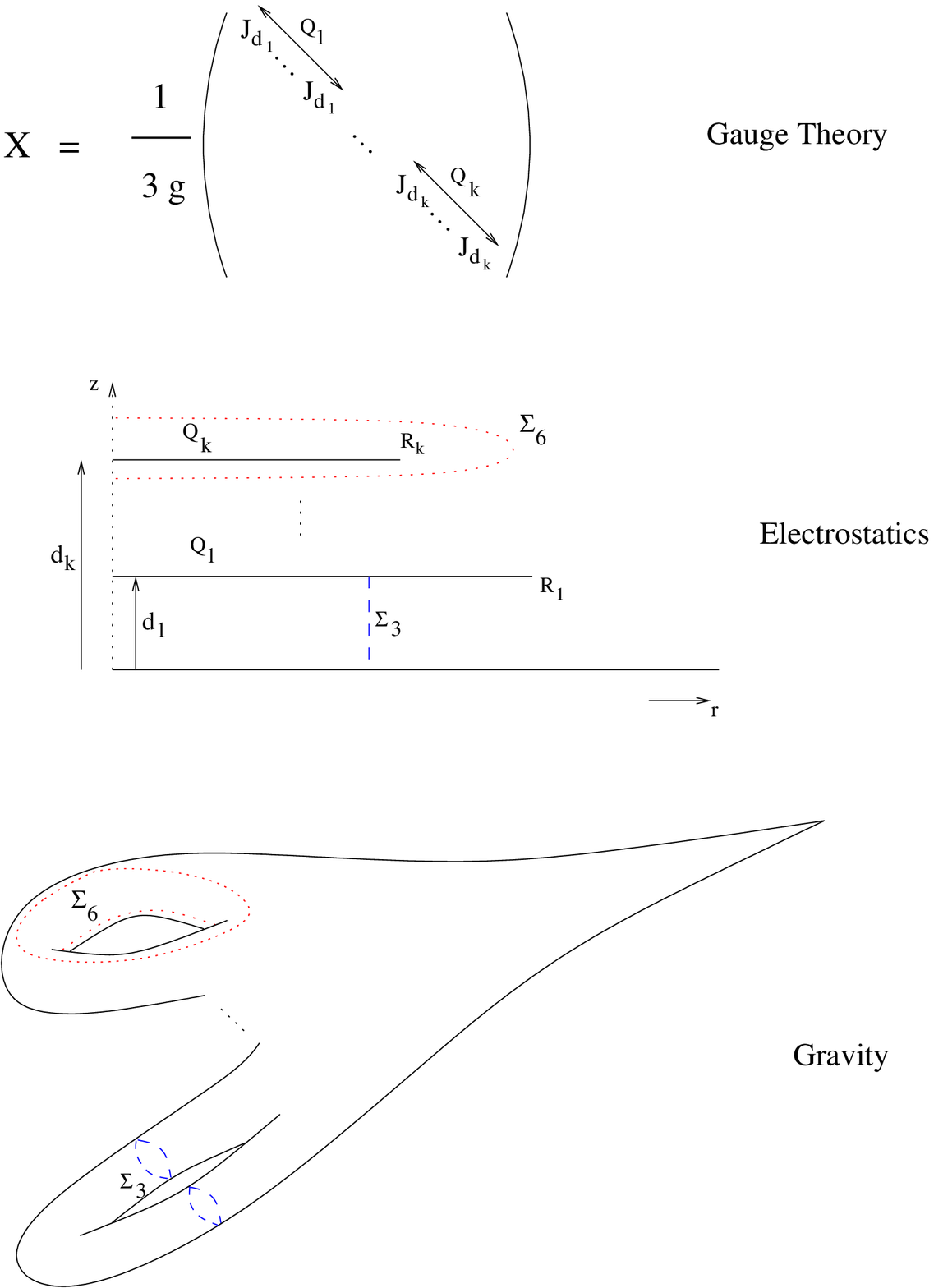}
 \caption{Mapping between matrix model vacua, electrostatics configurations, and geometries. For illustrative purposes, we have replaced the $S^2 \times S^5$s associated to each point $(r,z)$ with $S^0 \times S^0$. In the full geometry, the dotted segment maps to a submanifold $\Sigma_6$ that is topologically $S^6 \times S^2$ (simply connected) rather than the $S^1 \times S^0$ shown here. Similarly, the dashed segment maps to a submanifold $\Sigma_3$ that is topologically $S^5 \times S^3$ rather than the $S^0 \times S^1$ here.}
\end{figure}

\subsubsection*{Properties of the supergravity solutions}

We briefly review some properties of the supergravity solutions \cite{lm}. It is straightforward to show that as expected, all of these supergravity solutions approach asymptotically the near-horizon D0-brane solution. In the infrared region, the solutions have interesting topology, as we now recall. 

The coordinates $r$ and $z$ in the electrostatics problem form two of the nine spatial coordinates in the geometry. In addition, for each value of $r$ and $z$, we have an $S^2$ and an $S^5$ with varying radii. The $S^5$ shrinks to zero size on the $r=0$ axis, while the $S^2$ shrinks to zero size at the locations of the conducting plates, so we have various non-contractible $S^3$s and $S^6$s corresponding to paths that terminate on different plates or on different segments of the vertical axis respectively. This is illustrated in figure 1. As shown in \cite{lm}, through an $S^6$ corresponding to a path surrounding plates with a total charge of $Q$, we have $N_2 = 8Q/\pi^2$ units of flux from the dual of the Ramond-Ramond four-form, suggesting the presence of $N_2$ D2-branes. Similarly, through an $S^3$ corresponding to a path between plates separated by a distance $d$, we have $N_5 = 2 d /\pi$ units of H-flux, suggesting that this part of the geometry between the plates is describing the degrees of freedom of $N_5$ NS5-branes. 

If we take large plates at a fixed separation, the region between the plates corresponds to a long throat in the geometry with NS5-brane flux. Below, we will understand how to take a limit where such a throat becomes infinite so that we recover the Lin-Maldacena geometry.

\subsubsection*{Solution to the electrostatics problem (simplest case)}

We will now solve the electrostatics problem above in the simplest case of a single disk above the infinite plate, with the space in between corresponding to a single NS5-brane throat. Thus, we would like to find the potential for a conducting disk of charge $Q=\pi^2N_2/8$ a distance $d=\pi N_5 /2$ above the infinite conducting plate, in the presence of the background field (\ref{vinf}), as shown in figure 2. This will give the geometry dual to the matrix model vacuum with $N_2$ copies of the $N_5$ dimensional irreducible representation. 

\begin{figure}
\centering
\includegraphics{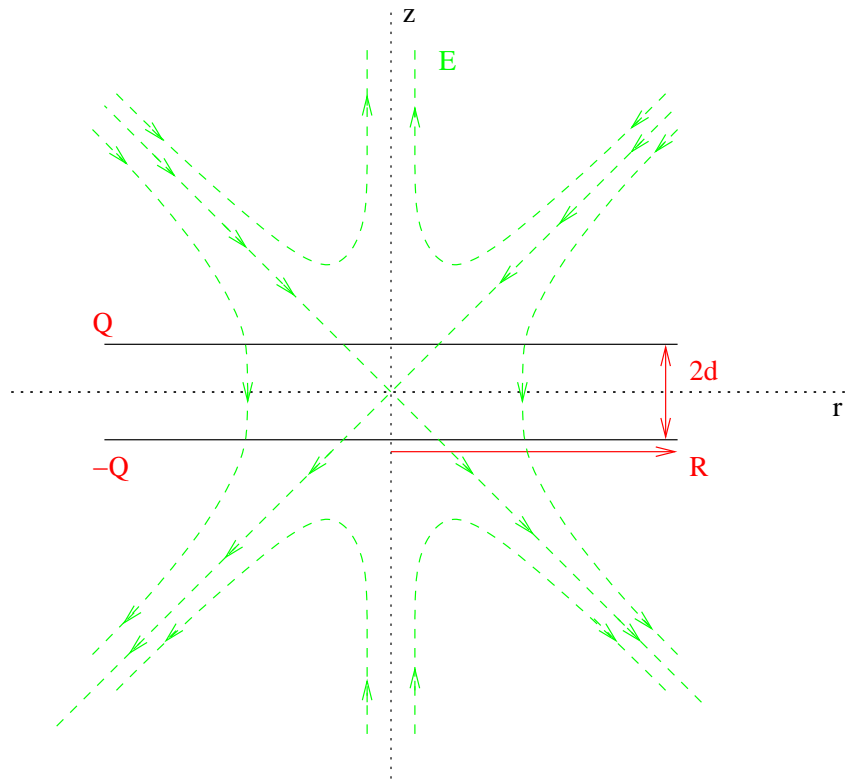}
 \caption{Electrostatics problem (cross section) corresponding to gravity solutions with single NS5 throat. The method of images has been used to replace the infinite conducting plate with an image disk.}
\end{figure}

We can simplify the problem somewhat using the the fact that electrostatics is linear and scale-invariant. Thus, if 
\[
V = (r^2z - {2 \over 3} z^3) + \phi_\kappa(r,z)
\]
is the solution to the electrostatics problem above with $V_0 = 1$, $R=1$, and $d=\kappa$ (with $Q=q(\kappa)$ determined by the condition of vanishing charge density at the edge of the disk), then the solution to the general problem will be
\be
\label{scaled}
V = V_0(r^2 z - {2 \over 3} z^3) +  R^3 V_0 \; \phi_{d \over R}(r/R,z/R) \; ,
\ee
and the charge on the disk will be
\be
\label{charge}
Q = q(d/R) V_0 R^4 \; .
\ee
Note that $\phi$ is the part of the potential that vanishes at infinity, arising from the charges on the disks. We would now like to determine $\phi_\kappa(r,z)$.

We may assume that the potential vanishes on the infinite conducting plate. Let us call the (constant) potential on disk $V=\Delta$. This will be determined in terms of $\kappa$ by the condition that the charge density vanishes at the edge of the disks, but for now, we will take it to be arbitrary and solve for the potential in general. 

By the method of images, the potential will be the same as for a pair of conducting disks at $z = \pm \kappa$ with potentials $V = \pm \Delta$, with the potential going like $V \to z r^2 - {2 \over 3} z^3$ at infinity. Without the background potential, this is a classic problem in electrostatics, considered by Maxwell, Claussius and Helmholtz, Kirchhoff, Polya and Szego, and eventually solved by Nicholson \cite{nicholson} and Love \cite{love}. For these references and a nice summary, see \cite{sneddon}. Fortunately, the method of solution may easily be extended to our case with the background potential, as we describe in appendix \ref{esapp}. 

To give the solution, it is convenient to define
\be
\label{defbeta}
\beta = \Delta + {2 \over 3} \kappa^3 \; .
\ee
Then the potential is given by 
\be
\label{basic}
\phi_\kappa(r,z) = {\beta \over \pi} \int_{-1}^1 G_\kappa(r,z,t) f_\kappa(t) dt \;  ,
\ee
where 
\[
G_\kappa(r,z,t) = - {1 \over \sqrt{r^2 + (z+\kappa+it)^2}} + {1 \over \sqrt{r^2 + (z-\kappa+it)^2}}
\]
and 
\be
\label{fdef}
f_\kappa(t) = f_\kappa^{(0)}(t) - 2 {\kappa \over \beta} f_\kappa^{(2)}(t) \; .
\ee
The functions $f_\kappa^{(n)}(t)$ are special functions  solving the integral equation
\be
f_\kappa^{(n)}(t) - \int_{-1}^1 K_\kappa(t,x) f_\kappa^{(n)}(x) = t^n
\label{fredholm}
\ee
with kernel
\[
K_\kappa(t,x) = {1 \over \pi} {2\kappa \over 4 \kappa^2 + (x-t)^2} \; . 
\]

This is a Fredholm integral equation of the second kind, and the solution may be written as a series
\be
\label{series}
f_\kappa^{(n)}(t) = \sum_{m=0}^\infty K_\kappa^m \circ t^n
\ee
where
\[
(K \circ g)(t) \equiv \int_{-1}^1 K(x,t) g(x) \; .
\]
It may be shown that the series converges for any value of $\kappa > 0$ to define a bounded continuous function $f_\kappa^{(n)}(t)$.

The function $f$ is related to the charge density on the disk as
\bea
f_\kappa(t) &=& {2 \pi \over  \beta} \int_t^1 {r \sigma_\kappa(r) dr \over (r^2-t^2)^{1 \over 2}} \cr
\sigma_\kappa(r) &=& { \beta \over \pi^2} \left[ {f_\kappa(1) \over (1-r^2)^{1 \over 2}} - \int_r^1 {f_\kappa'(t)dt \over (t^2-r^2)^{1 \over 2}}\right] \; ,
\label{density}
\eea
so that the total charge on the disk is 
\be
\label{chargefromf}
q(\kappa) = {\beta \over \pi} \int_{-1}^1 f_\kappa(t) dt \; .
\ee
 
\subsubsection*{Vanishing charge density constraint} 

The solution in the previous section was for arbitrary potential $\Delta$, and will generally have a charge density that is nonvanishing at the tip of the disk. We will now determine a formula for $\Delta(\kappa)$ for which the charge density vanishes.

\begin{figure} \label{qfig}
%GNUPLOT: LaTeX picture with Postscript
\begin{picture}(0,0)%
\includegraphics{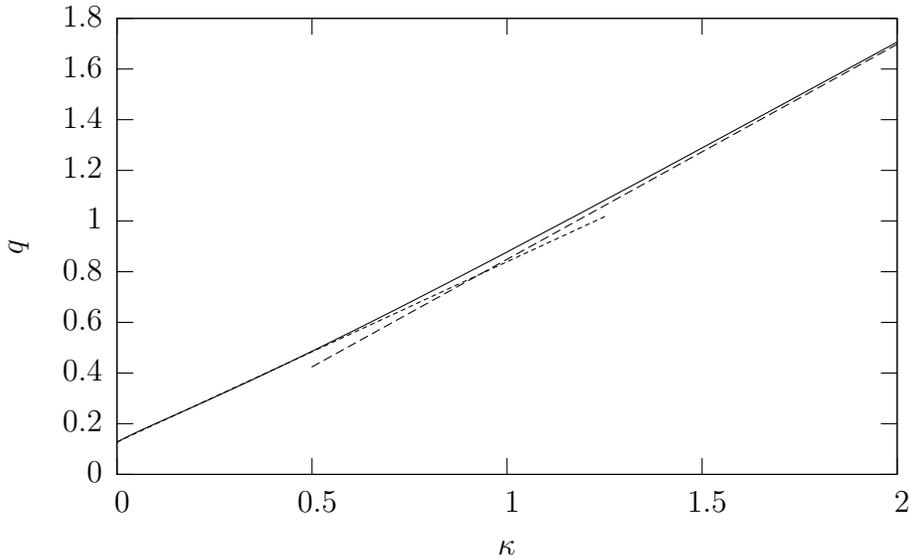}%
\end{picture}%
\begingroup
\setlength{\unitlength}{0.0200bp}%
\begin{picture}(18000,10800)(0,0)%
\put(2200,1650){\makebox(0,0)[r]{\strut{} 0}}%
\put(2200,2606){\makebox(0,0)[r]{\strut{} 0.2}}%
\put(2200,3561){\makebox(0,0)[r]{\strut{} 0.4}}%
\put(2200,4517){\makebox(0,0)[r]{\strut{} 0.6}}%
\put(2200,5472){\makebox(0,0)[r]{\strut{} 0.8}}%
\put(2200,6428){\makebox(0,0)[r]{\strut{} 1}}%
\put(2200,7383){\makebox(0,0)[r]{\strut{} 1.2}}%
\put(2200,8339){\makebox(0,0)[r]{\strut{} 1.4}}%
\put(2200,9294){\makebox(0,0)[r]{\strut{} 1.6}}%
\put(2200,10250){\makebox(0,0)[r]{\strut{} 1.8}}%
\put(2475,1100){\makebox(0,0){\strut{} 0}}%
\put(6150,1100){\makebox(0,0){\strut{} 0.5}}%
\put(9825,1100){\makebox(0,0){\strut{} 1}}%
\put(13500,1100){\makebox(0,0){\strut{} 1.5}}%
\put(17175,1100){\makebox(0,0){\strut{} 2}}%
\put(550,5950){\rotatebox{90}{\makebox(0,0){\strut{}$q$}}}%
\put(9825,275){\makebox(0,0){\strut{}$\kappa$}}%
\end{picture}%
\endgroup
\caption{Plot of $q(\kappa)$. Dashed lines are the asymptotes $\frac{8}{3\pi}\kappa$ for large $\kappa$ and $1/8 + 0.711 \kappa$ for small $\kappa$.}
\end{figure}

From (\ref{density}), it follows that $\sigma(1)=0$ if and only if $f(1)=0$ and $f'(1)$ is bounded. From the series solution, it is straightforward to prove that the latter condition is always satisfied for the functions $f_n(t)$, so our constraint comes from requiring $f(1)=0$. Now, from (\ref{fdef}) and the definition (\ref{defbeta}) of $\beta$, we see that the condition $f(1)=0$ determines $\Delta$ to be
\be
\label{givesdelta}
\Delta(\kappa) =  2 \kappa {f^{(2)}_\kappa(1) \over f^{(0)}_\kappa(1)} - {2 \over 3} \kappa^3 \; .
\ee
Finally, the charge on the disk is given in terms of $\kappa$ by
\be
\label{givesq}
q(\kappa)=  {f^{(2)}_\kappa(1) \over f^{(0)}_\kappa(1)} \; {2 \kappa \over \pi} \int_{-1}^1 f^{(0)}_\kappa(t) dt - {2\kappa \over  \pi} \int_{-1}^1 f^{(2)}_\kappa(t) dt \; .
\ee

Via equation (\ref{charge}) this function $q$ determines the radius of the disk in the original problem in terms of the charge $Q$, the potential $V_0$, and the separation $d$. The function $q$ is plotted in figure \ref{qfig}. We show in appendix \ref{limapp} that its limiting behavior for small and large $\kappa$ is
\be
\ba{llll}
q(\kappa) &\to &{1 \over 8} \qquad \qquad &\kappa \to 0 \; , \cr 
q(\kappa) &\to &{8 \over 3 \pi} \kappa \qquad \qquad &\kappa \to \infty.
\ea
\ee
We will see in section \ref{nBPSsec} that the function $q(\kappa)$ is physically important since it computes the energies of certain near-BPS states in the theory.

\subsubsection*{Summary}

In summary, to generate the supergravity solution dual to the vacuum of the plane-wave matrix model corresponding to $N_2$ copies of the $N_5$ dimensional irreducible representation, we:
\begin{itemize}
\item choose $\kappa$ (ultimately related to a choice of coupling $g$), determine $\phi_\kappa(r,z)$ from (\ref{basic}) and $q(\kappa)$ from equation (\ref{givesq})
\item take $R = (\pi N_5)/(2 \kappa)$ so that $d = R \kappa = \pi N_5 / 2$ 
\item choose $V_0 = (2 \kappa^4 N_2)/(q(\kappa) \pi^2 N_5^4)$ so that (using (\ref{charge})) $Q = q(\kappa) V_0 R^4 = \pi^2 N_2/8$
\end{itemize}   
Then the gravity dual is given by equations \eqref{IIA ansatz last} with $V$ given by \eqref{scaled}.

\section{Matching Parameters with Gauge Theory}

In discussing the various scaling limits of the theory, we will need to understand 
how the gauge theory parameters match with the parameters in the supergravity solution, or
equivalently, with the parameters $d$, $R$, $Q$, and $V_0$ in the electrostatics problem. As we mentioned, the parameters $d$ and $Q$ are proportional to the number of units of NS5-brane and D2-brane flux through the noncontractible $S^3$ and $S^6$ in the geometry. As shown in \cite{lm}, this allows us to associate \footnote{Note that we have taken all quantities in the electrostatics problem to be dimensionless.}
\[
d = {\pi \over 2} N_5 
\]
and 
\[
Q = {\pi^2 \over 8} N_2 \; .
\]

The remaining electrostatics parameter is $V_0$ (or equivalently, $R$), which we interpret in the gauge theory as follows. The asymptotic form of the geometry is determined by the background field in combination with the leading dipole fields arising from the charges on the plates. These depend respectively only on $V_0$ and the combination $dQ$, which is proportional to the dipole moment. These asymptotics should be the same for all vacua of a given theory, so the parameters $V_0$ and $dQ$ must depend only on $g$ and $N=N_2 N_5$, the parameters that determine which matrix model we are talking about. This is clearly true for $dQ$, but we must also have
\be
\label{fact1}
V_0 = f_1(g^2, N_2 N_5) \; .
\ee

To further constrain $V_0$ it is useful to note that in the gauge theory, the planar amplitudes depend only on $N_5$ and the combination $g^2 N_2$. This suggests that the free string 
theory on the dual spacetime should be controlled by these two parameters, and in particular, these parameters should control the metric. In terms of the electrostatics parameters, it is straightforward to see that the metric depends only on $d$ and $R$ (scaling $V_0$ while holding these fixed scales the dilaton and Ramond-Ramond fields, but leaves the metric fixed), so $d$ and $R$ should each be some function of $N_5$ and $g^2 N_2$. This is clearly true for $d$, but we must also have
\be
\label{fact2}
R = f_2(g^2 N_2, N_5) \; .
\ee
Using the relation $Q = V_0 R^4 q(d/R)$, together with (\ref{fact1}) and (\ref{fact2}) we may conclude that
\be
\label{vzero}
V_0 = {1 \over g^2} h(g^2 N_2 N_5)
\ee
for some function $h$. 

In section \ref{D2sec}, we will see that the D2-brane limit of the gauge theory discussed in section \ref{gaugesec} matches with the corresponding limit of the supergravity solution only if the function $h$ approaches some constant $h_\infty$ at large values of its argument. Since we will mostly be interested in this regime ($g^2 N_2 N_5$ is always large when supergravity is valid), we get the identification
\be
\label{hlim}
V_0 = {h_\infty \over g^2} \; .
\ee
The electrostatics parameter $R$ is a more complicated function of the gauge theory parameters, but follows from the other identifications via (\ref{charge}). 

\section{Scaling limits}

In this section, we consider various scaling limits of the gravity theory in which one of the three parameters is scaled to infinity, with the others scaled such that we end up with something nontrivial. 

\subsection{Large $d$: the D2-brane limit} \label{D2sec}

We begin by considering a limit of large $d$ with fixed $R$. To understand how we should scale $V_0$ to leave us with a nontrivial supergravity solution, note that for large $d/R$, the formula (\ref{charge}) determining the charge on the disks becomes
\be
\label{newcharge}
Q = {8 \over 3 \pi} V_0 d R^3 \; ,
\ee
where we have used the large $\kappa$ behavior of $q(\kappa)$. Also, the potential \eqref{vinf}, taken near the position of the disk by replacing $z = d + \eta$, becomes
\be
\label{newpotl}
V = -{2 \over 3} V_0 d^3 - 2 V_0 d^2 \eta  + V_0 d (r^2 - 2 \eta^2) + V_0 (\eta r^2 - {2 \over 3} \eta^3) \; .
\ee
The first two terms here have no effect on the supergravity solution, since the solution (\ref{IIA ansatz last}) depends only on $\partial_z^2 V$ and $\partial_r V$. Thus, from (\ref{newpotl}) and (\ref{newcharge}), we see that in order to leave a finite nontrivial background potential and a finite non-zero charge on the disk, we must take $V_0$ to scale like $1/d$. Thus, our limit is 
\be
\label{D2limit}
d \to \infty \qquad \qquad Q \; \; {\rm fixed} \qquad \qquad V_0 d = W_0 \; \; {\rm fixed} \; . 
\ee
In this limit, we have a single charged conducting disk (with no infinite plate) in a background potential 
\be
\label{newback}
V = W_0 (r^2 - 2\eta^2) \; .
\ee

In \cite{lm} Maldacena and Lin wrote down explicitly the geometry corresponding to this situation. In appendix \ref{D2app}, we give an alternate derivation of the solution by explicitly solving the electrostatics problem, verifying that (\ref{newcharge}) correctly gives the charge necessary to ensure that the charge density vanishes at the edge of the plates. For large $r$, the solution approaches the solution for near-horizon D2-branes but with the flat directions along the D2-branes replaced by an $S^2$ \cite{lm}. Thus, we expect that this limit should correspond to the limit of the matrix model giving rise to D2-branes on $S^2$. Using the correspondence between matrix model parameters and electrostatics parameters (in particular, assuming that the function $h$ in (\ref{vzero}) is simply a constant at large argument), we find that the limit (\ref{D2limit}) becomes, 
\[
N_5 \to \infty \qquad \qquad N_2 \; \; {\rm fixed} \qquad \qquad {g^2 \over N_5} \; \; {\rm fixed},
\]
which is precisely the D2-brane limit discussed in section \ref{gaugesec}. 

\subsection{Large $V_0$: the 't Hooft limit}

The next limit we consider is the 't Hooft limit discussed in section \ref{gaugesec},
\[
N_2 \to \infty \qquad \qquad N_5 \; \; {\rm fixed} \qquad \qquad {g^2 N_2} \; \; {\rm fixed},
\]
 which appeared to be a decoupling limit retaining interacting fivebrane degrees of freedom. Using the correspondence between matrix model parameters and electrostatics parameters, we find that this is a limit with
\[
Q \to \infty \qquad \qquad d \; \; {\rm fixed} \qquad \qquad R \; \; {\rm fixed} \qquad \qquad V_0 \to \infty
\]
From the supergravity point of view, this is a limit in which the metric is held fixed with the maximum value of the dilaton going to zero. Thus, we have free string theory on the background corresponding to a single finite-sized disk above the infinite conducting plate.

This geometry contains a finite throat region with NS5-brane flux, but also a noncontractible $S^3$ with D2-brane flux. Thus, while we are describing fivebrane degrees of freedom, this limit of the gauge theory does not correspond to the infinite-throat Lin-Maldacena solution for NS5-branes on $S^5$.

\subsection{Large $R$: Little string theory on $S^5$}

Finally, we would like to understand precisely what limit of our solution is required to obtain the Lin-Maldacena infinite-throat solution for NS5-branes on $S^5$. This corresponds to an electrostatics problem with two infinite conducting plates, with the potential between the plates equal to 
\be
\label{infplates}
V = {1 \over g_0} \, \sin \left( {\pi z \over d} \right) \, I_0 \left( {\pi r \over d} \right).
\ee
To obtain this from our solution, we certainly need to take a limit where $R$ is going to infinity with $d$ fixed. However, generically, we would simply end up with a constant vertical electric field between the plates. This does not give rise to any metric (the supergravity fields depend only on $\partial_z^2 V$ and $\partial_r V$) so we must take $V_0$ large enough so that the leading corrections to this constant electric field remain nonzero in the limit. 

To understand the proper scaling, we start by considering the electrostatics solution for finite $R$. In the $r<R$ region between the plates, we have an axially symmetric solution to the Laplace equation regular at $r=0$, so we can write 
\[
V(r,z) = V_{z=d} {z \over d} + \sum_{n=1}^\infty c_n \sin \left( {n \pi z \over d} \right) \, I_0 \left( {n \pi r \over d} \right) \; .
\] 
Here, we have separated off a constant electric field term that does not affect the metric such that the remaining piece vanishes at $z=0$ and $z=d$ for $r<R$. Now the $c_n$s are determined by the potential at $r=R$, 
\be
\label{limint}
c_n =  \left(I_0 \left( {n \pi R \over d} \right)\right)^{-1} V_0 R^3 {2 \over d} \, \int_0^d dz \, \left({1 \over R^3} (R^2 z  - {2 \over 3} z^3) + \phi_{d \over R} (1, {z \over R} )- \Delta_{d \over R} {z \over d} \right) \, \sin \left( {n \pi z \over d} \right)
\ee
where we have used the expression (\ref{scaled}) for $V$. Using our solution for $\phi$, it is simple to show that the integral here has only a power law dependence on $R$, so the large $R$ behavior of $c_n$ is dominated by the exponential damping coming from the Bessel function at large argument,  
\[
(I_0(z))^{-1} \sim {  \sqrt{2 \pi z} e^{-z}} \; . 
\]
To compensate for this damping, we must scale $V_0$ exponentially in $R$,
\be
\label{vscale}
V_0 \sim e^{\pi R \over d},
\ee
which allows us to keep $c_1$ finite in the limit. All the other coefficients $c_{n>1}$ still vanish in the limit, so we indeed end up with the Lin-Maldacena solution. To be more precise, we can evaluate the integral in (\ref{limint}) to find the prefactor in (\ref{vscale}). Using the results of appendix C.1, we find the behavior
\be
\label{estimate}
c_1 \to V_0 \, C \, (Rd)^{3 \over 2} \, e^{-\pi R \over d} 
\ee
where we have numerically estimated $C$ to be $C \approx 0.080$. Thus, the precise limit we need to take to recover (\ref{infplates}) is  
\[
R \to \infty \qquad \qquad d \; \; {\rm fixed} \qquad \qquad V_0 \to {1 \over g_0} \, {1 \over C} \, (Rd)^{-{3 \over 2}} \, e^{\pi R \over d} \; ,
\]
which also implies $Q \to \infty$.

In supergravity language, the limit we are taking is designed to take the NS5-brane throat infinite while holding the dilaton at the bottom of the throat fixed. The fact that V is exponentially damped as we go towards the middle of the plates gives rise to the linear dilaton behavior of the final supergravity solution.
  
\subsubsection*{Field theory description} 

Using the correspondence of parameters between field theory and the electrostatics we find that the limit of the plane-wave matrix model that defines the dual of the Lin-Maldacena solution, i.e. the field theory description of Little String Theory on $S^5$, is 
\be
\label{fixed}
N_2 \to \infty \qquad \qquad N_5 \; \; {\rm fixed} \qquad \qquad {1 \over g^2} \lambda^{3 \over 8} e^{-a\lambda^{1 \over 4} /N_5} \; \; {\rm fixed} \; .
\ee 
where $a$ is a numerical coefficient related to the constant in (\ref{hlim}) by $a = 2(\pi^2/h_\infty)^{1 \over 4}$. Thus, rather than holding the 't Hooft coupling fixed, we scale it to infinity in a controlled way 
\be
\label{lambdascaling}
\lambda \sim N_5^4 \ln^4(N_2) \; . 
\ee
Unfortunately, this implies that perturbation theory is not useful on the field theory side, though perhaps there are some near-BPS sectors of the theory where the expansion parameter is not the naive 't Hooft coupling.

\section{Discussion}

\subsection{The double-scaling limit}

We have seen that to obtain the field theory dual of the Lin-Maldacena supergravity solution for Little String Theory on $S^5$, we need to take a large $N_2$ limit while scaling the 't Hooft coupling to infinity in a controlled way. 
This double scaling limit is reminiscent of limits used to define low-dimensional string theories in old matrix models (see for example \cite{gm}). There, the 't Hooft coupling is scaled to some critical value in a controlled way as $N$ goes to infinity, such that all terms in the genus expansion continue to contribute even though $N$ becomes infinite. We suspect that this is also the situation here, except that in our case, the ``critical'' value of the 't Hooft coupling is infinity. The fact that $N_2$ becomes infinite is consistent with the fact that the dilaton vanishes asymptotically in the supergravity solution. On the other hand, we still have a string genus expansion in the bulk of the supergravity solution, so we can understand the scaling of $\lambda$ to infinity as necessary for the gauge theory to reproduce nontrivial string interactions in the bulk.

To understand this in more detail, consider the matrix model genus expansion for some physical observable. It takes the form
\be
\label{genusexp}
F = \sum_n {f_n(g^2 N_2, N_5) \over N_2^n}
\ee
where $f_n(\lambda, N_5)$ gives the sum of genus $n$ diagrams. In the 't Hooft limit with $\lambda = g^2 N_2$ fixed and $N_2$ taken to infinity, only the planar $n=0$ term contributes. What we are suggesting is that the scaling (\ref{lambdascaling}) is such that all terms in the expansion (\ref{genusexp}) contribute. If this is true, it predicts that the large $\lambda$ behavior of $f_n$ is 
\be
\label{predict}
f_n(\lambda) \to a_n \left[ \lambda^{5 \over 8} e^{a \lambda^{1 \over 4} \over N_5} \right]^n \; .
\ee
The quantity in square brackets divided by $N_2$, which we can call $\tilde{g}$, is the inverse of the quantity being held fixed in (\ref{fixed}), so the genus expansion (\ref{genusexp}) becomes
\[
F = \sum_n a_n \tilde{g}^n \; ,
\]
Thus, the constant $\tilde{g}$ serves as the effective string coupling.  

The behavior (\ref{predict}) is a nontrivial prediction of our results and the assumption that the string theory genus expansion is still related to the gauge theory genus expansion. It should apply to the behavior of any physical observable that survives the scaling limit, for example the energy of any state in the matrix model corresponding to some excitation in the NS5-brane throat. It would be interesting to understand more precisely from the gauge theory point of view which set of observables remain in the limit. 

Unfortunately, it seems difficult to check the behavior (\ref{predict}) directly from the gauge theory, since it would involve summing infinite sets of diagrams. However, it may be that this is possible for certain BPS or near-BPS observables, as for the circular Wilson loop in ${\cal N}=4$ supersymmetric Yang-Mills theory, where for example the full contribution at the planar level is given by an infinte set of ladder diagrams that can be summed explicitly \cite{esz}. Intriguingly, that result
\[
\langle W \rangle_{N= \infty} = \sqrt{2 \over \pi} \lambda^{-{3 \over 4}} e^{\sqrt{\lambda}} 
\]
takes a rather similar form to our prediction here.\footnote{Note that the $\sqrt{\lambda}$ in this example and the $\lambda^{1 \over 4}$ in our case both represent the squared radius of the respective $S^5$s in string units.} A specific limit in which some matching similar to \cite{bmn} might be possible is in a Penrose limit of the geometry \cite{nspp}, associated with geodesics around the $S^2$ at $r=0$. 

\subsection{An infinite parameter family of NS5-brane solutions}

Starting from our exact supergravity solution for the simplest class of vacua, we have found a specific limit that gives the Lin-Maldacena solution corresponding to the region between two infinite conducting plates. It is interesting to note that this solution is actually the simplest in an infinite-parameter family of solutions. In the electrostatics language, we can have any function
\[
V = \sum_{n=1}^\infty c_n \sin \left( {n \pi z \over d} \right) \; I_0 \left( {n \pi r \over d} \right)
\]
with $c_n$ chosen to fall off fast enough so that the sum converges for all $r$. The supergravity solution corresponding to any such potential will have an infinite throat with noncontractible $S^3$ carrying fivebrane flux. 

While it does not seem possible to obtain these more general solutions as limits of the solution we considered in this paper, it is plausible that we could obtain them as limits of solutions corresponding to more general vacua of the plane-wave matrix model. Specifically, we could imagine starting with a solution containing some arbitrarily large number of disks, taking the size of the lowest disk to infinity as before, but now tuning all of the parameters $V_0$, $d_2$, $Q_2$, ... , $d_n$, $Q_n$ in such a way that all of the upper disks have some non-trivial influence on the potential between the plates. It would be interesting to understand better the physical interpretation of these more general solutions.

\subsection{Energies of near-BPS states} \label{nBPSsec}

A useful result coming from our solution is the formula (\ref{charge}) that determines the radius of the disk in terms of the other parameters $d$, $Q$, and $V_0$. From the supergravity solution (\ref{IIA ansatz last}), it is straightforward to show \cite{lm} that $R$ determines the radius of the $S^5$ at the point $(r=R,z=d)$ corresponding to the edge of the disk as\footnote{To see this, we use the Laplace equation to rewrite $V''$ and note that $\partial_r V$ vanishes on the disks.}
\[
R^2_{S^5} / \alpha' = 4 R .
\]
As described in \cite{lm} section 2.2, this in turn determines the energies of certain near-BPS states with large angular momentum on the $S^5$ (see equation (2.42)).\footnote{These are the states of string theory on a plane wave obtained by a Penrose limit associated with geodesics going around the $S^5$ at fixed $r$ and $z$.} Thus, our function $q(\kappa)$ in (\ref{charge}), defined in (\ref{givesq}) and plotted in figure 3, determines the near-BPS energies for large $R$ and $d$ but arbitrary $d/R$, interpolating between the small $d/R$ and large $d/R$ results given in \cite{lm}, equations (2.84) and (2.57) respectively. 

\subsection{Other definitions of Little String Theory on $S^5$}

Finally, we note that while the limit we have defined may be the simplest description of the Little String Theory on $S^5$, there should be many other field theoretic definitions. In the electrostatics picture, it should arise any time two nearby disks are scaled to infinite radius at fixed separation with the background potential scaled so that the potential between the plates remains nontrivial. Thus, we could start with more general vacua of the plane wave matrix model, vacua of the D2-brane field theory on $S^2$, or vacua of ${\cal N} = 4$ SYM on $R \times S^3/ Z_k$.\footnote{This latter possibility was noted in \cite{lm}} It would be interesting to understand more precisely the field theory limits for these other cases also. 

\section*{Acknowledgements}

We would like to thank Allan Adams, Ofer Aharony, Joel Feldman, Richard Froese, Shiraz Minwalla, Sunil Mukhi, Joe Polchinski, Gordon Semenoff, Eva Silverstein, Stephen Shenker, Herman Verlinde, Wan Yan Wong and especially Emil Martinec for discussions. We would also like to thank Tom Davis and Dastegir Al-Quaderi for discussions of numerical issues. We would like to thank the Banff International Research Station and the Weizmann Institute of Science where parts of this work were completed. This work has been supported in part by the Natural Sciences and Engineering Research Council of Canada, the Killam Trusts, the Alfred P. Sloan Foundation, and the Canada Research Chairs programme.

\appendix

\section{Supergravity solutions} \label{sgapp}

The general Lin-Maldacena $SU(2|4)$-symmetric supergravity ansatz (suppressing an overall factor of $\alpha'$ in the metric) is given by \cite{lm} 
\begin{eqnarray} 
ds_{10}^{2} &=& \left( \ddot V - 2 \dot V \over - V'' \right)^{1/2}
\left\{ - 4 { \ddot V \over \ddot V - 2 \dot V } dt^2 + { - 2 V'' \over \dot V} ( d\rho^2
+ d\eta^2 )  + 4 d\Omega_5^2 + 2{ V'' \dot V \over   \Delta }d\Omega_2^2
\right\}  \cr
e^{4\Phi  } &=&{ 4 ( \ddot V - 2 \dot V)^3 \over - V'' \dot V^2
\Delta^2 }  \cr
C_{1} &=&- { 2 \dot V' \, \dot V \over \ddot V - 2 \dot V } dt \label{IIA ansatz last} \\
F_{4} &=&d C_3  ,\quad \quad \quad
 \quad \quad C_{3}=- 4  { \dot V^2 V'' \over   \Delta } dt\wedge d^{2}\Omega , \cr
H_{3} &=&d B_2 ~,~~~~~~~~~~~~~B_{2}   = 2 \left( { \dot V \dot V' \over   \Delta} + \eta \right) d^{2}\Omega 
\cr
  \Delta &\equiv& (\ddot V - 2 \dot V) V'' - ( \dot V')^2
\nonumber  
\end{eqnarray}
Their explicit solution corresponding to Little String Theory on $S^5$ is 
\begin{eqnarray}
 ds_{10}^{2} &=&N_5 \left[ -2r \sqrt{\frac{I_{0}}{I_{2}}}dt^{2}+2r \sqrt{%
\frac{I_{2}}{I_{0}}}d\Omega _{5}^{2}+\sqrt{\frac{I_{2}}{I_{0}}}\frac{I_{0}}{%
I_{1}}(dr ^{2}+d\theta ^{2})+\sqrt{\frac{I_{2}}{I_{0}}}\frac{%
I_{0}I_{1}s^{2}}{I_{0}I_{2}s^{2}+I_{1}^{2}c^{2}}d{\Omega }_{2}^{2}\right]  \notag
\\ \label{ns5sol}
B_{2} &=& N_5 \left[ \frac{-I_{1}^{2}cs}{I_{0}I_{2}s^{2}+I_{1}^{2}c^{2}}+\theta \right]
 d^{2}\Omega  \\
e^{\Phi } &=& g_0 N_5^{3/2} 2^{-1}\left( \frac{I_{2}%
}{I_{0}}\right) ^{\frac{3}{4}}\left( \frac{I_{0}}{I_{1}}\right) ^{\frac{1}{2}%
}\left( I_{0}I_{2}s^{2}+I_{1}^{2}c^{2}\right) ^{-\frac{1}{2}} \\
C_{1} &=&- g_0^{-1} N_5^{-1} \frac{4 {I_{1}}^2 c}{I_{2}}dt \\
C_{3} &=&- g_0^{-1} \frac{4I_{0}I_{1}^{2}s^{3}}{I_{0}I_{2}s^{2}+I_{1}^{2}c^{2}}%
dt\wedge d^{2}\Omega  
\end{eqnarray}
where $g_0$ is a constant, $I_n(r)$ are the usual Bessel functions, $s \equiv \sin(\theta)$, and $c \equiv \cos(\theta)$.

\section{Matrix model conventions} \label{mapp}

In the usual conventions, the plane-wave matrix model is described by a Hamiltonian
\bea
\label{PPmatrix}
H &=& M_p^2 R \tr \left( {1 \over 2} P_A^2 - {1 \over 4} [X_A, X_B]^2
- {1 \over 2} \Psi^\top \gamma^A [X_A, \Psi] \right)\cr
&& + {M_p^2 R \over 2} \tr \Big( {\left({\mu\over
 R M_p^2}\right)^2 } (X_i/3)^2 +  \left({\mu \over R M_p^2}\right)^2 (X_a/6)^2 
\cr
&& \qquad \qquad + {i \over 4} \left( {\mu \over R M_p^2} \right) \Psi^\top \gamma^{123} \Psi
+ {2i \over 3} \left( {\mu \over R M_p^2} \right) \epsilon^{ijk} X_i X_j X_k \Big)\ .
\eea
With these conventions, the large $N$ limit with $\mu$ fixed and $N/R$ fixed describes the sector of
M-theory on the plane-wave background with 
\[
P^+ = {N \over R} 
\] 
and $P^-$ identified with $H$. In the conventions of this paper, we define
\[
g^2 = \left( {M_p^2 R \over \mu} \right)^3
\]
and we take
\[
\tilde{X} = g^{-\frac{1}{3}} X \qquad \qquad \tilde{H} = {1 \over \mu} H ,
\]
where the tilded quantities are the ones used in the rest of this paper.

\section{Solution of the electrostatics problem} \label{esapp}

In this appendix, we describe the solution of the electrostatics problem described in section \ref{gsec}, namely to find a solution of Laplace's equation
\[
\nabla^2 V = 0
\]
such that $V=0$ at $z=0$, $V=\Delta$ for $\{z=\kappa,0<r<1\}$ and $V$ behaves as $z r^2 - {2 \over 3} z^3$ for large $z^2+r^2$.

To start, we write
\[
V = z r^2 - {2 \over 3} z^3 + \phi 
\]
so that $\phi$ vanishes at infinity and at $z=0$ and is given by
\[
\phi(r) = \Delta + {2 \over 3} \kappa^3 - \kappa r^2 \equiv \beta(1-\alpha r^2)
\]
on the upper plate. Axially symmetric solutions to the Laplace equation that are regular at $r=0$ and vanish as $r \to \infty$ are linear combinations of functions
\[
e^{\pm zu} J_0 (ru),
\]
where $u$ is a continuous parameter. If we denote by $V_+$ and $V_0$ the function $\phi$ in the regions $z \ge d$ and $0 \le z \le d$ respectively, then we must have \bea
V_0 &=& \int_0^\infty B(u) \sinh(zu) J_0(ru) du \cr
V_+ &=& \int_0^\infty C(u) e^{-zu} J_0(ru) du \label{vsoln}
\eea
taking into account the boundary conditions at $z=0$ and $z=\infty$. 

We now take into account the boundary conditions at $z=\kappa$. First, we must have $V_0=V_+$ at $z=\kappa$, so it must be that
\beas
B(u) &=& 2 \beta u^{-1} e^{-\kappa u} A(u) \cr
C(u) &=& 2 \beta u^{-1} \sinh(\kappa u) A(u)  
\eeas
for some function $A(u)$ (the prefactors have been chosen for later convenience). Also, we have $\partial_z V_0 = \partial_z V_+$ at $z=\kappa$ for $r>1$ and $V_0 = V_+ = \beta(1-\alpha r^2)$ for $0 < r <1$. These will be satisfied if
\bea
\int_0^\infty A(u) J_0(r u) du &=& 0 \qquad \qquad \qquad \; \; \; r > 1 \label{int1}\\ 
\int_0^\infty u^{-1} (1-e^{-2 \kappa u}) A(u) J_0(r u) du &=& 1 - \alpha r^2 \qquad \qquad 0 < r < 1 \label{int2} .
\eea
This set of equations is of the type considered in \cite{sneddon}, chapter 4.6, ``Dual integral equations with Hankel kernel and arbitrary weight function''. For convenience, we review the solution of this type of equation in appendix \ref{dualeq}. For our specific case, the solution is given by
\[
A(u) = {2u \over \pi} \int_0^1 f(t) \cos(ut) dt \; ,
\]
where $f(t)$ is an even function of $t$ satisfying the integral equation
\be
\label{inteq}
f(x) - \int_{-1}^1 K(x,t) f(t) dt = g(x) \; ,
\ee
where 
\be
\label{givesg}
g(x) = 1 - 2 \alpha x^2
\ee
and 
\[
K(x,t) = {1 \over \pi} {2 \kappa \over 4\kappa^2 + (x-t)^2} \; .
\]
The solution $f$ of the integral equation may be given as a series (\ref{series}). Substituting our solution for $A$ in (\ref{vsoln}) gives the simpler result (\ref{basic}), valid for all $z$.

\subsection{Limiting forms of the solution} \label{limapp}

In our discussion of scaling limits of the theory, it will be useful to have some explicit results for the behavior of the electrostatics solution for large and small values of $\kappa$.   

\subsubsection*{Large $\kappa$}

For large $\kappa$, the kernel $K$ has a small norm, so the series solution (\ref{series}) is well approximated by its leading term. Thus, 
\[
f(x) \to g(x) = 1 - 2 \alpha x^2 \; .
\]
In order that the charge density vanishes at the edge of the plate, equation (\ref{density}) implies that $f(1)$ must vanish. Thus, we must have
\[
\alpha = {1 \over 2} \; ,
\]
which (for large $\kappa$) implies
\[
\Delta \to -{2 \over 3} \kappa^3 + 2 \kappa \; .
\]
This implies that $\beta = 2 \kappa$, so that using (\ref{chargefromf}), we have
\[
q \to {8 \over 3 \pi} \kappa \; .
\] 
As a check, we can match these results with those in appendix \ref{D2app}, where we have solved the electrostatics problem after first taking the limit of large $d$ with fixed $R$. 

\subsubsection*{Small $\kappa$}

For small $\kappa$, the problem is more difficult to study, since the norm of the kernel $K$ approaches one (the kernel approaches a delta function), so the series solution (\ref{series}) converges very slowly. The work \cite{hutson} studied in detail the case of charged conducting disks at small separation without a background potential, but some of the general discussion there is helpful in our case also. 

Up to corrections of (fractional) order $\sqrt{\kappa}$, the leading small $\kappa$ behavior of the solution of the integral equation (\ref{inteq}) is \cite{hutson} 
\[
f(t) \to \kappa^{-1} \int^1_{-1} k(s,t) g(s) ds \; ,
\]
where
\[
k(s,t) = {1 \over 2 \pi} \log \left\{ {1 -st + (1-s^2)^{1 \over 2} (1-t^2)^{1 \over 2} \over 1 -st - (1-s^2)^{1 \over 2} (1-t^2)^{1 \over 2} } \right\} \; .
\]
This corresponds to the approximation that $\sigma(r)$ varies as $\kappa^{-1} \phi(r)$.

In our case, $g$ is given by (\ref{givesg}), so we find that 
\be
\label{leadingf}
f(t) \to {1 \over 2 \kappa} \left\{ (1-t^2)^{1 \over 2} - \alpha (1 - t^2)^{1 \over 2} + {2 \over 3} \alpha (1-t^2)^{3 \over 2} \right\} \; .
\ee
Now, in order that $\sigma$ vanishes at the edge of the disk, (\ref{density}) implies that $f(1)$ vanishes and that the $f'(1)$ is bounded. We see that the latter condition is satisfied only if $\alpha \to 1$ in this limit.\footnote{For finite values of $\kappa$, the series solution (\ref{series}) may be used to show that $f'(1)$ is actually bounded for all $t$. On the other hand, corrections to the formula (\ref{leadingf}) are generally non-zero at $t = \pm 1$. Thus, the vanishing charge density constraint for the exact solution at finite $\kappa$ comes from the condition $f(1)=0$. We have verified numerically in appendix \ref{numerics} that this condition gives the same result $\alpha \to 1$ in the limit $\kappa \to 0$ as the condition that $f'(1)$ is bounded applied to the leading approximation.} Thus, in the limit of small $\kappa$, we have
\be
\label{smallkappa}
f^\kappa(t) \to {1 \over 3 \kappa} (1-t^2)^{3 \over 2}\; .
\ee
From $\alpha \to 1$, we infer that 
\be
\label{dsk}
\Delta \to \kappa
\ee
in the limit, and using (\ref{chargefromf}) we have 
\[
q(\kappa) \to {1 \over 8} \; .
\]

We are particularly interested in the small $\kappa$ behavior of the integral (\ref{limint}), which after a change of variables becomes
\be \label{integral}
I = \int_0^1 dy \sin (\pi y) \left[ \phi_\kappa(1, \kappa y) - \left\{\Delta_\kappa y - \kappa y + {2 \over 3} \kappa^3 y^3 \right\} \right]  \; .
\ee
The largest terms are the first two terms in curly brackets, each of which give a contribution of order $\kappa$, but (\ref{dsk}) implies that these cancel. Using numerical methods described in appendix \ref{numerics}, we have estimated the remaining contributions. We find that there is a further cancelation between the $\phi$ term and the term in curly backets, both of which give contributions of order $\kappa^2 \ln(\kappa)$. The net result is that the integral behaves for small $\kappa $ as\footnote{Note that here we cannot simply use the leading approximation (\ref{smallkappa}) for $f$. The reason is that the integral over $t$ receives most of its contribution close to the boundaries of the interval $[-1,1]$ where the leading approximation goes to zero like a $3/2$ power. Corrections to (\ref{smallkappa}), which are fractionally small in the bulk of the interval, become more important in the region near the boundaries, and result in a modified behavior for the integral.}  
\[
I \approx 0.040 \kappa^2 .
\]

While this result is numerical, we have performed an analytic check on the method. If we normalize the function
\[
\tilde{\phi}(x,y) = C \left[ \phi_\kappa(1+x, \kappa y) - \left\{\Delta_\kappa y - \kappa y + {2 \over 3} \kappa^3 y^3 \right\} \right]
\]
by choosing $C$ such that (for example) $\tilde{\phi}(0,0.5) = 1$, then $\tilde{\phi}$ should have a well definied limit as $\kappa \to 0$. This must be a nontrivial solution to the Laplace equation in the case where we have one infinite conducting plate (at $y=0$), and one semi-infinite conducting plate (at $y=1$, $x<0$), with vanishing potential on both plates (recall that the term invoving $\Delta$ ensures that $\tilde{\phi}$ vanishes on both plates). This reduces to a two-dimensional problem that may be solved using conformal mapping techniques \cite{lm}. As explained in \cite{lm}, if we define a complex coordinate $z=x+iy$ and another complex variable
\[
w = 2 \partial_z V = \partial_x V - i \partial_y V \; .
\] 
Then the Laplace equation ensures that the mapping between $z$ and $w$ is analytic. Further, $\partial_x V$ is everywhere nonnegative and vanishes at the plates, so the region outside the plates must map to the right half plane, with the plates mapping to the imaginary axis. The explicit transformation that acheives this (unique up to transformtions generated by translations, scalings, and inversions that fix the imaginary axis) is \cite{lm}
\be
\label{defw}
z = i w + {1 \over \pi} \ln(w) + {i \over 2} + {1 \over \pi} + {1 \over \pi} \ln(\pi) \; .
\ee
This implies
\[
V = Re \left( \int^z w(z') dz' \right) = u\left({1 \over \pi} -v\right) \; ,
\]
where $w=u+iv$. From (\ref{defw}), we find that $u$ and $v$ are determined in terms of $x$ and $y$ by
\beas
x &=& -v + {1 \over 2\pi} \ln(u^2 + v^2) + {1 \over \pi} + {1 \over \pi} \ln (\pi) \cr
y &=& u + {1 \over \pi} \tan^{-1} \left( {v \over u} \right) + {1 \over 2}  \; ,
\eeas
so we have a relatively explicit result for the potential in the limit. After normalizing the result in the same way that we normalized $\tilde{\phi}$, we find excellent agreement with our numerical results for $\tilde{\phi}$ in the limit of small $\kappa$.

\section{Single disk solution} \label{D2app}

In this appendix, we solve the electrostatics problem obtained in the limit of large $d$ with $V_0 d=W_0$ fixed and $R$ fixed, to give a more direct derivation of the asymptotically near-horizon D2-brane solution derived by Lin and Maldacena. 

To define this solution, we want to find the potential for a charged conducting disk of radius $R$ and charge $Q$ in a background potential
\be
\label{newback2}
V = W_0 (r^2 - 2\eta^2) \; .
\ee
In this case, the potential $W_0$ will eventually be related to the charge, $Q$, and the radius, $R$, by the condition that the charge density vanishes at the edge of the disk.

By the superposition and scale-invariance properties of electrostatics, the solution must take the form
\[
V = W_0 R^2 \phi(r/R,\eta/R),
\]
where $\phi$ is the solution to the problem with $W_0=1$ and $R=1$. Further if $q$ is the total charge on the disk required in this simplified problem so that the charge density vanishes at the edge of the disk, we must have
\[
Q = W_0 R^3 q .
\]

To start, we would like to find
a solution to Laplace's equation with boundary conditions that the
potential is fixed to $\Delta$ on a disk of radius 1 at $z=0$ and 
becomes $r^2 - 2 z^2$ at infinity. The potential, $\Delta$, and the
total charge on the disk, $q$, will be fixed by demanding that the
charge density vanishes at the edges of the disk. 

We begin by writing 
\[
V = r^2 - 2 z^2 + \phi \; ,
\]
such that $\phi$ vanishes at infinity and 
\[
\phi(r,z=0) = \Delta - r^2 ;\ .
\]
If we denote by $V_+$ the potential $\phi$ for $z>0$, then separating variables gives 
\be
\label{Vbess}
V_+(r,z) = \int_0^\infty u^{-1} A(u) e^{-uz} J_0(ur) du \; .
\ee
By symmetry, the potential $V_-$ for $z<0$ must be
\[
V_-(r,z) = V_+(r,-z) \; .
\]
Finally, we require that 
\beas
V_+ = V_- &=& \Delta - r^2 \qquad \qquad 0 \le r \le 1 \; , \cr
\partial_z V_+ - \partial_z V_- &=& 0 \qquad \qquad \qquad \;  \; r > 1.
\eeas
These imply the dual integral equations
\bea
\int_0^\infty A(u) J_0(ur) du &=& 0 \qquad \qquad \qquad \; \; \; r > 1 \\ 
\int_0^\infty u^{-1} A(u) J_0(ur) du &=& \Delta - r^2 \qquad \qquad 0 < r < 1, 
\eea
which are of the type considered in appendix \ref{dualeq}, with $k(u) = 0$ and $h(r) = \Delta - r^2$. In this case, the integral equation is trivial, so the result is
\[
A(u) = {2u \over \pi} \int_0^1 f(t) \cos(ut) dt
\] 
with 
\[
f(t) = g(t) = \Delta - 2 r^2 \; .
\]
Substituting for $A$ in (\ref{Vbess}) gives 
\[
V_+ = {1 \over \pi} \int_{-1}^1 {f(t) \over \sqrt{r^2 + (z+it)^2}}  dt
\]
The charge density on the disk is given by
\beas
\sigma(r) &=& {1 \over 4 \pi} (\partial_z V_-(r,-\epsilon) - \partial_z V_+(r,\epsilon)) \cr
&=&{1 \over 2 \pi} \int_0^\infty A(u) J_0 (ur) du \cr
&=& {1 \over \pi^2} \left[{f(1) \over (1-r^2)^{1 \over 2}} - \int_r^1 ds {f'(s) \over (s^2 - r^2)^{1 \over 2}} \right] \cr
&=& {\Delta - 2 \over \pi^2} {1 \over \sqrt{1 - r^2}} + {4 \over \pi^2} \sqrt{1-r^2} \; .
\eeas
In order that the charge density vanishes on the tip, we need 
\[
\Delta = 2 \; ,
\]
so finally
\[
\sigma(r) = {4 \over \pi^2} \sqrt{1-r^2}
\]
and the total charge is 
\[
q = {8 \over 3 \pi} \; .
\]
If we like, we can write an explicit solution for the potential using oblate spherical coordinates (see \cite{sneddon}, section 3.3), but we will not need it here. 

\section{Dual integral equations}\label{dualeq}

In this appendix, we review the solution of dual integral equations of the form
\bea
\int_0^\infty A(u) J_0(ru) du &=& 0 \qquad \qquad \qquad \; \; \; r > 1 \\ 
\int_0^\infty u^{-1} (1+k(u)) A(u) J_0(ru) du &=& h(r) \qquad \qquad 0 < r < 1\; ,
\eea
following chapter 4.6 of \cite{sneddon}. In general, the solution is given as
\[
A(u) = {2u \over \pi} \int_0^1 f(t) \cos(ut) dt \; ,
\]
where $f(t)$ is the solution to a Fredholm integral equation of the second kind,
\[
f(x) + \int_{0}^1 K(x,t) f(t) = g(x) \;,
\]
with 
\[
g(x) = {d \over dx} \int_0^x {u h(u) du \over \sqrt{x^2-u^2}}. 
\] 
The kernel $K$ is given in terms of $k(u)$ by
\[
K(x,u) = {1 \over \sqrt{2 \pi}} \{K_c(|x-u|) + K_c(x+u)\} \; ,
\]
with
\[
K_c(\xi) = \sqrt{2 \over \pi} \int_0^\infty k(t) \cos(\xi t) dt \; .
\]

\section{Numerical calculations} \label{numerics}

In this appendix we outline the numerical methods used to solve the integral
equation (\ref{fredholm}). We solved this equation using the
Nystr\"om method (e.g.~\cite{numbook}). This method consists of discretizing the
interval and integrating using numerical quadrature. To ensure the validity of
our numerical results we checked that the number of points in the
discretization was sufficiently large so that increasing it did not affect the
results, and by checking that the matrix that must be inverted in this method
is not nearly singular. The Nystr\"om method has been applied before
to integral equations similar to (\ref{fredholm}) in fluid mechanics
\cite{fluid}. The equation (\ref{fredholm}) with $n=1$ arises in \cite{fluid}
as a limiting case in the analysis of waves radiated from an oscillating
submerged disk. Although in \cite{fluid} Gauss-Legendre quadrature was used, we
found that in our case the solution was accurately produced by simply using the
midpoint rule. By extrapolating the numerical solution we found that the ratio
$f^{(0)}_\kappa/f^{(2)}_\kappa \to 2$ as $\kappa \to 0$. Using these results we
were also able to estimate
\[
\Delta(\kappa) -\kappa + \frac23\kappa^3=  2 \kappa \left({f^{(2)}_\kappa(1) \over f^{(0)}_\kappa(1)} - {1 \over 2}\right) ,
\]
finding that for small $\kappa$ this had the form $-0.7\kappa^2\ln(\kappa)$.
We obtained an approximate form for $q(\kappa)$ by doing a straightforward
numerical integral of the solution. Similarly, we obtained an estimate of
(\ref{integral})
by doing a numerical integral of the solution to the integral equation to find
the potential, $\phi$, and then performing the integral in (\ref{integral})
numerically.

\end{document}